\RequirePackage[T1]{fontenc}
\documentclass[lettersize,journal]{IEEEtran}
\usepackage{amsmath,amsfonts}
\usepackage{algorithmic}
\usepackage{algorithm}
\usepackage{array}
\usepackage{textcomp}
\usepackage{stfloats}
\usepackage{url}
\usepackage{verbatim}
\usepackage{graphicx}
\usepackage{svg}
\usepackage{cite}
\usepackage{multirow}
\usepackage{multicol}
\usepackage{subcaption}
\usepackage{subcaption}
\usepackage[skip=5pt]{caption} 
\usepackage{caption}  
\usepackage{booktabs} 
\usepackage{hyperref}
\usepackage{booktabs}
\usepackage[table]{xcolor}

\begin{document}

\title{A Rate-Distortion Optimized Skip Mode Coding Framework for Neural Video Compression in AVS}

\author{Yuntian Jiang, Shuai Li,~\IEEEmembership{Senior Member,~IEEE}, Yanbo Gao, Jinglin Zhang, Hui Yuan,~\IEEEmembership{Senior Member,~IEEE}, \\Mao Ye,~\IEEEmembership{Senior Member,~IEEE}
\thanks{Manuscript received Oct. 19, 2025.}
\thanks{This work was supported in part by the National Natural Science Foundation of China (No. U24B20132, 62572275 and 62271290 ), and the Shandong Provincial Natural Science Foundation (No. ZR2024LZN021, ZR2025QB44 and ZR2025MS1074). }
\thanks{Y. Jiang, S. Li, J. Zhang and H. Yuan  are with School of Control Science and Engineering, Shandong University, Jinan 250100, China. 
\par Y. Gao is with School of Software, Shandong University, Jinan, China, and also with Shandong University-WeiHai Research Institute of Industrial Technology, Weihai 264209, China.  E-mail:  ybgao@sdu.edu.cn.
\par M. Ye is with University of Electronic Science and Technology of China, Sichuan, China. } 
}



\maketitle

\begin{abstract}
With the development of deep learning, neural video compression (NVC) is gaining increasing interest and is currently under standardization in the AVS (Audio Video Coding Standard). 
Skip mode coding, a crucial component in the traditional video compression, is also adopted in AVS NVC. 
Although the skip mode takes a large portion in encoding the motion and residual features in the NVC, there is a lack of investigation on the rate distortion performance of the skip mode coding. 
In this paper, we first thoroughly investigate the rate and distortion distribution in the skip mode coding within a spatial frame and under different temporal reference qualities, and demonstrate that distortion distribution changes spatially and temporally. 
Therefore, a rate-distortion optimized skip mode coding (RDOSkip) framework is proposed, where the rate and distortion in the skip mode coding is formulated based on a statistical distribution predicted by the hyperprior and the Bayes' theorem. 
An adaptive skip mode decision method based on spatial-temporal joint RD optimization is developed for the skip mode coding among channels and frames in a group of pictures. 
It addresses the limitation of the rigid fixed-distortion distribution assumption in skip mode coding within the existing methods that rely on a fixed skip threshold. 
The temporal skip mode decision method has been adopted in the ongoing AVS NVC standard and incorporated into the reference software. 
{The proposed RDOSkip can be directly used in the inference without any training and with negligible complexity increase.}
Experimental results further demonstrate that the proposed RDOSkip can improve the performance by 0.90\% and 1.29\% on AVS and HEVC datasets, respectively, in terms of BD-rate saving.
\end{abstract}


\section{Introduction}

\begin{figure}[!t]
\centering
\setlength{\belowcaptionskip}{-15pt} 
\includegraphics[width=\columnwidth]{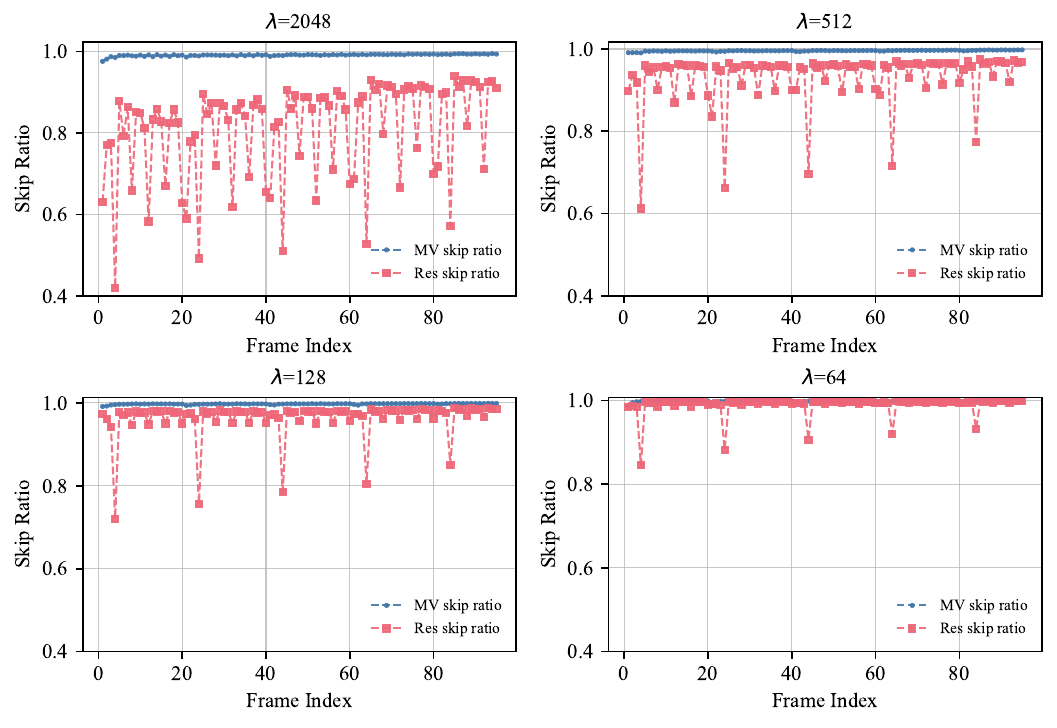}
\caption{The skip ratio of motion and residual features, on the BQTerrace sequence. The skip ratio refers to the percentage of skipped elements in the latent feature using the AVS-EEM.}
\label{intro_1}
\end{figure}

\IEEEPARstart{V}{ideo} compression is a fundamental task in video transmission. Traditional video coding standards, such as Advanced Video Coding (AVC)\cite{264}, High Efficiency Video Coding (HEVC)\cite{265} and Versatile Video Coding (VVC)\cite{266}, have achieved remarkable success through decades of development. In recent years, with the development of deep learning, neural video compression (NVC)\cite{DCVC-DC,TSAN, DCVC-HTM,MPAI_EEV,Deep_affine, Ensemble_RDO_E2E, Hign_efficiency_iamge, Machine_perception,Rethinking_latent} has emerged as a promising alternative, showing continuous improvement. It is also under standardization in the AVS (Audio Video Coding Standard)\cite{fan2004overview, yu2009overview,wang2009robust,zhang2009context,zhang2019recent,M8819,M8973}. Among their key components, skip mode is used in the traditional methods\cite{264,265,266} and neural video compression methods\cite{DCVC-RT,Alphavc}. It compresses coding units or features with negligible residuals by representation with inter prediction alone (without coding the residual), effectively reducing the bitrate without significantly increasing the distortion. It is known that skip mode takes a large portion of the coding modes at inter-frame coding, i.e., over 70\%\cite{Skip_ratio} in the traditional video compression at certain quality parameter settings. In Fig.~\ref{intro_1}, we further illustrate the percentages of the skip modes used in the AVS NVC codec (noted as AVS-EEM). The detailed settings are further described in the experimental setting and the same setting is used for the following results. For NVC, the skip mode is referred to the skip mode coding of motion vector features and residual features. It can be seen that the percentage of skip mode coding for both motion vector features and residual features are very high (all over 40\%), especially at lower rates (over 90\%), as shown in Fig. 1. Therefore, the coding efficiency of skip mode is very important in the overall performance of NVC. Moreover, the percentage of skip mode coding for motion vector features are much higher than that of residual features, generally over 96\% and even over 99\% in certain video frames. Considering that most of the motion vector features are skipped, in this paper, we focus on investigating the skip mode coding of residual features, and it is simply referred to skip mode unless otherwise noted. 

In the traditional video coding methods, skip mode decisions are made based on rate-distortion optimization (RDO)\cite{skip_264_1,skip_264_2,skip_264_3,li_lagrangian}, which compares the RD cost of skip mode against the other coding modes. Consequently, it can be considered as RD optimal at the coding of each coding unit. 
The skip mode is indicated by a skip flag which is transmitted to the decoder.
By contrast, NVC methods commonly adopt a variational auto-encoder (VAE) architecture\cite{VAE,VAE_2020}, 
incorporating analysis and synthesis transforms\cite{balle2016end,GaGs_0,GaGs_1,AINN}, motion estimation and compensation\cite{DMVC,ECVC,PRA}, quantization and entropy modeling\cite{balle_Hyper,mix_gau, Gen_gaussian}, and the skip mode is generally determined in the entropy coding process without RDO.
Specifically, the existing NVC methods usually adopt a hyperprior based entropy coding, where hyper-encoder and hyper-decoder branches are used to predict the mean and variance of the latent feature, in order to construct a probabilistic model for entropy coding. 
Skip mode\cite{DCVC-RT,Alphavc} is then determined based on the predicted variance by comparing it with a fixed threshold, to skip the encoding of the latent feature and replace it with the predicted mean.  This can help reduce bitrate under low-variance conditions that indicate a high confidence in the predicted mean. 
Instead of transmitting a skip flag as in the traditional video coding, a binary mask is generated at both the encoder and decoder by the comparison process.

\begin{figure*}[!t]
\centering
  \captionsetup[subfigure]{skip=2pt}
  \captionsetup[subfigure]{labelfont={color=black}}
 \subfloat[]{\includegraphics[width=0.44\textwidth]{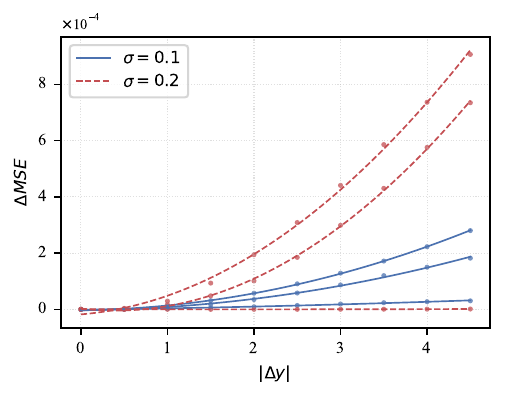}%
\label{intro2a}}
\hfill
\captionsetup[subfigure]{labelfont={color=black}}
\subfloat[]{\includegraphics[width=0.44\textwidth]{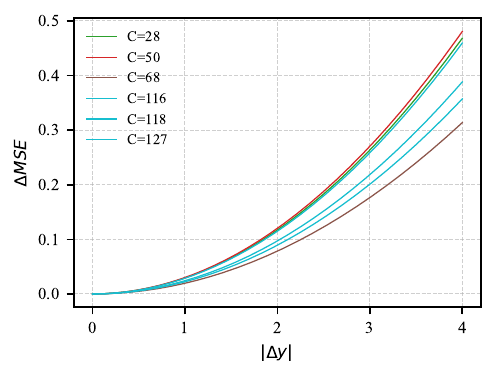}%
\label{intro2b}}
\caption{Illustration of the MSE variation in the image domain corresponding to feature loss $|\Delta y|$ in the feature domain, on the BQTerrace, $\lambda=2048$. (a) for different feature elements under different variances, 
and (b) for different channels. (a) illustrates that feature elements with the same variance can also lead to different MSE losses while (b) illustrates different channels exhibit distinct sensitivities to changes in $\Delta y$, 
leading to varying magnitudes of distortion responses.
}
\label{intro_2}
\end{figure*}

Considering that the bitrate of each feature element is modeled using a Gaussian or Laplace entropy model, the estimated rate for each variance is the same for all feature elements with a fixed feature deviation $|\Delta y|$ (the absolute difference between the feature element value and the predicted mean). However, distortion may vary significantly across feature elements. As shown in Fig.~\ref{intro2a}, distortions corresponding to different $|\Delta y|$ are illustrated at different feature elements under two variances. Here, features are directly added with different values to mimic the different feature deviations $|\Delta y|$. Fig.~\ref{intro2a} demonstrates that under the same variance, distortion does not vary as consistently as rate and even changes significantly over different variances. Moreover, distortions of different channels, instead of distortion per feature element for statistical robustness, are also illustrated as shown in Fig.~\ref{intro2b}. It can be seen that the distortion also exhibits significant variation across different feature channels. In this way, the skip mode, determined using a fixed threshold of the variance, can be considered as rate-based optimization and lacks explicit consideration of distortion. Moreover, the large variations in terms of the distortion suggest that a unified skip threshold is suboptimal, and a spatially varying skip mode decision is highly desired for NVC.

\begin{figure}[!t]
\centering
\setlength{\belowcaptionskip}{-15pt} 
\includegraphics[width=0.9\columnwidth]{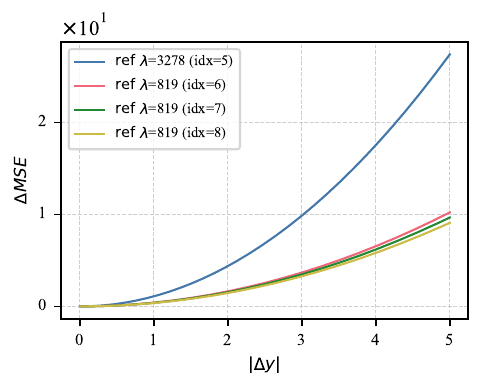}
\caption{The impact of the current frame with different qualities of the reference frame, on BQTerrace, $\lambda=2048$. Idx is the index of the current frame in the sequence. The ref $\lambda$ denotes the quality of the reference frame. The MSE loss is calculated frame-wise.}
\label{intro_3}
\end{figure}

In addition, as shown in Fig.~\ref{intro_1}, the skip ratio of the residual feature follows a hierarchical pattern within a Group of Pictures (GOP) which is highly consistent with the hierarchical quality structure used in the AVS NVC model. This indicates that the skip mode coding may also be affected temporally by the quality of the reference frame due to inter prediction. Fig.~\ref{intro_3} further illustrates the distortion distribution under $|\Delta y|$ when the reference frames are of different qualities. It can be seen that the distortion distribution varies significantly under different reference frame qualities. This also changes the rate-distortion optimization for skip mode decisions and indicates a temporal quality (or Lagrange multiplier $\lambda$) dependent skip mode decision. In a word, a rate-distortion optimized skip mode decision based on different spatial and temporal distortion properties is highly desired. 

In this work, we aim to formulate the rate-distortion optimization process for the skip mode coding in neural video compression (NVC). Especially for the distortion in the reconstruction, since it is measured in the pixel domain while skip mode decisions are made in the latent feature domain, it cannot be directly obtained. In this paper, we bridge this gap through a Bayesian-based distortion formulation that couples the pixel-domain distortion with the feature-domain loss. Together with the rate model based on the entropy, a spatially RD optimized skip mode decision method is developed. Moreover, in view of the GOP-based hierarchical quality structure, a temporally dependent skip mode decision method is further proposed to take the reference frame quality fluctuation into consideration. 

The main contributions of this paper can be summarized as follows:

\begin{itemize}
    \item[  1)] The rate and distortion optimization process of skip mode coding is first investigated for the NVC framework. The distortion of the reconstruction in the pixel domain is formulated via Bayes' theorem, to construct a mapping to the feature loss in the feature domain. To the best of our knowledge, this is the first work that investigates the RDO for skip mode coding in NVC.
    \item[  2)] A spatially channel-wise RD optimized skip mode decision method is developed to facilitate the adaptive skip mode coding among channels. It adapts the variance threshold used in the skip mode decision based on the different distortion distributions corresponding to the feature loss.
    \item[  3)] A temporally RD optimized skip mode decision method is further proposed to explore the distortion change due to the reference frame quality fluctuation. The effect of the context-based encoder and decoder is investigated in the skip mode coding and hierarchical variance thresholds are determined based on the hierarchical quality structure in a GOP.
\end{itemize}

Notably, our method can be directly applied in the inference without any training. Experimental results demonstrate that the proposed method achieves BD-Rate savings of around 1.29\% on the HEVC dataset and 0.90\% on the AVS test dataset. Part of the proposed method has been adopted in the ongoing AVS NVC standard as M8789\cite{M8789} and incorporated into the reference software.

\section{Related Work}
\subsection{Skip Mode Coding in the Traditional Video Compression}
Skip mode coding is an important coding tool in the traditional video compression for inter-frame coding\cite{skip_264_1,skip_265_1}. It skips the entropy coding of the motion and residuals in the coding units when the residuals are relatively negligible compared with the bits used to encode the whole coding unit. In the traditional video encoding, each coding block has multiple candidate encoding modes, including the skip mode. The encoder decides the encoding mode for each coding unit by calculating and comparing the RD cost for each mode. Skip mode takes a large portion in the inter prediction modes. Therefore, to reduce computation brought by the full RD cost comparison over all the modes, fast skip mode decision methods have been proposed. They generally utilize temporal and spatial contextual information to predict the RD loss of skipping a block early. In \cite{Fskip_0}, a low-complexity skip mode prediction method was proposed, which predicts the RD cost of skip mode by referring the skip mode decision of the previous frame. In addition to skip mode where motion vectors and residual are both skipped, merge mode is also used in the traditional video compression where only the motion vectors are skipped and replaced with predicted motion vectors in the merge list. Both skip mode and merge mode are optimized via RD cost comparison to achieve the optimal coding performance.  

\subsection{Skip Mode Coding in the Neural Video Compression}
With the development of deep learning, neural video compression\cite{DVC,FVC,luguo,NVC_early_0,NVC_early_1,NVC_early_2}, has attracted lots of interests. DVC\cite{DVC} replaces traditional coding modules with neural networks, including motion estimation/compensation, transformation, entropy coding, etc., and uses RD loss for training, where the rate is approximated with the entropy. FVC\cite{FVC} proposed motion estimation and motion alignment in the feature domain with deformable convolution. DCVC\cite{DCVC,DCVC-TCM,DCVC-FM,DCVC-CM} proposed the context-based encoder and decoder based on the perspective that conditional entropy modeling of latent features is no larger than the traditional residual-based entropy. Unlike traditional codecs that compute residuals via direct pixel-wise subtraction, the context based encoder and decoder in DCVC uses learnable residual modeling, transforming residual computation to a conditional processing based on convolution. The subsequent works of DCVC have gradually surpassed traditional codecs in terms of performance and improved the efficiency of encoding and decoding. There are also some entropy coding methods\cite{checkboard_entropy,learning_switchable} based on hyperprior, spatial and channel context to enhance the entropy coding performance.

The ongoing NVC standard in the AVS (Audio Video Coding Standard) also adopts the context based encoder and decoder architecture. Optical flow\cite{Hyper_video,MMVC} based motion estimation\cite{M8967} and compensation is used to obtain the inter prediction as the context. 
Hyperprior based entropy coding\cite{mix_gau,M9028} is applied for the optical flow and residual feature coding, where a Gaussian based probabilistic model is formulated with mean and variance predicted by the hyperprior. Optimization of entropy models is one of the major development directions in the field of compression. Quantization is used before entropy coding, where uniform noise is added during training and rounding is used at the test as in\cite{balle2016end}. The rate distortion loss is used in the same way as the existing NVC methods, where multiple models are trained based on different Lagrange multipliers to achieve models for coding at different qualities. Moreover, a group of frames(GoP) based local Lagrange multiplier adaptation strategy is adopted as in\cite{DCVC-RT,DCVC-FM}, to achieve a hierarchical quality structure. 

Skip mode coding has been used in most of the existing NVC methods including DCVC\cite{DCVC-RT}, and AVS\cite{M7881,Alphavc}. It skips the entropy coding of the motion vector and residual features by comparing the variance predicted by the hyperprior with a predefined threshold. AlphaVC\cite{Alphavc} formulated the skip mode coding in a probabilistic manner based solely on the estimated rate. While it offers an efficient way of reducing bitrate, it fails to incorporate the effect of the distortion due to the skip mode coding, leading to suboptimal compression performance. In a word, while skip mode coding is used in most of the NVC methods, there is a lack of investigation on the full rate distortion optimization of the skip mode coding. This paper fills this gap by developing a rate distortion optimized skip mode coding method for the NVC framework, which adjusts the skip mode decision based on both the predicted rate and distortion to achieve RD optimal coding performance. 

\begin{figure*}[!t]
\centering
\includegraphics[width=\textwidth]{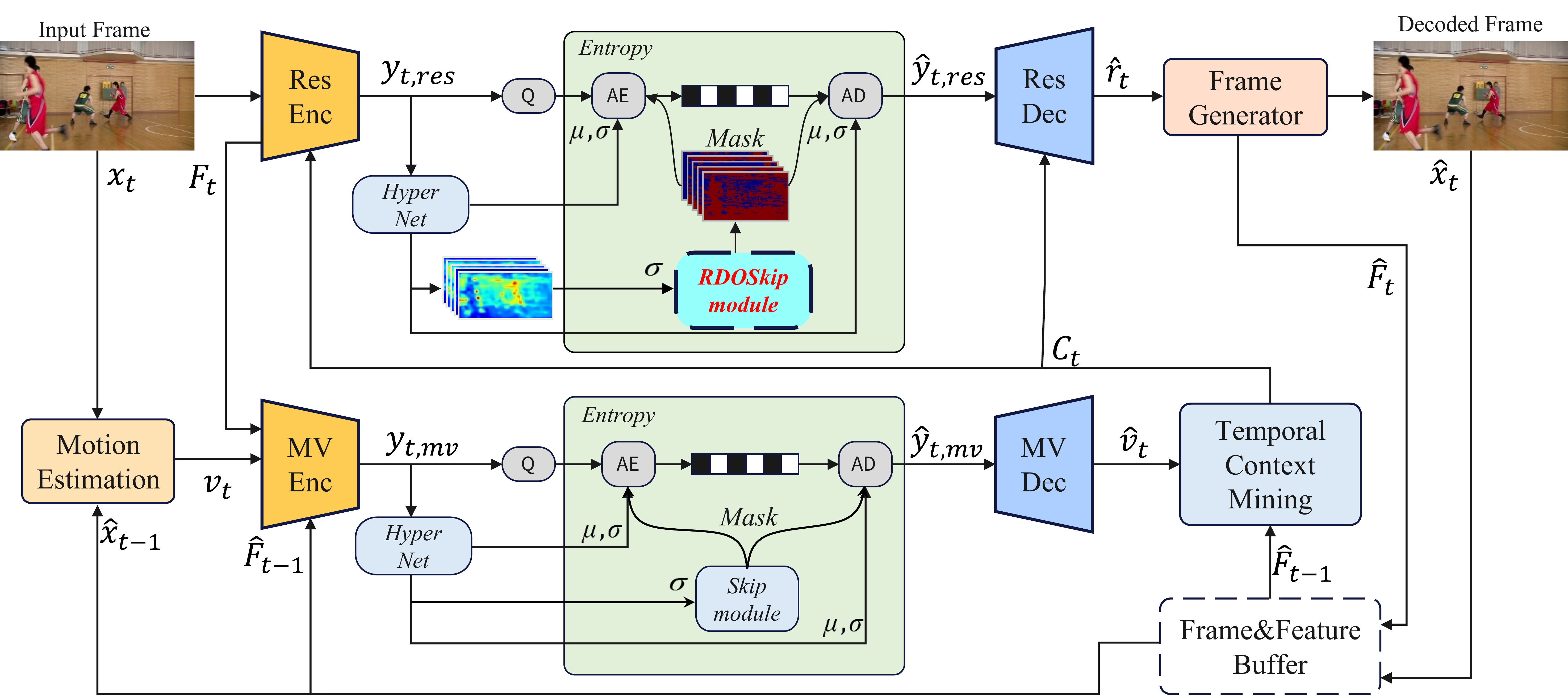}
\caption{Overview of the neural video compression based on our proposed adaptive skip module.}
\label{structure1}
 \end{figure*}

\section{Proposed Method}
\subsection{Skip Mode Coding Formulation in the Spatial Domain}
The framework of the current AVS Neural Video Compression (NVC) is illustrated in Fig.~\ref{structure1}, which adopts a similar structure to existing NVC methods. It includes an encoder/decoder (also known as the residual encoder and frame generator), motion estimation and compensation modules, hyperprior-based entropy encoders/decoders for both motion vector features and residual features, and quantization. The hyperprior-based entropy encoder/decoder further contains an entropy coding module, that encodes the hyperprior into the bitstream. The decoded hyperprior is then used to predict the mean and variance of residual feature for entropy formulation. The components in NVC are optimized with the rate-distortion loss using end-to-end learning. Skip mode coding is adopted in the hyperprior-based entropy coding for both motion vector and residual features, with a generated binary mask indicating whether skipping the coding of each feature element in the feature. As noted in the Introduction, we here only discuss the skip mode coding of the residual features, since most of the motion vector features are skipped and show less influence on the coding performance.

When the skip mode is used, the feature element in the residual feature is not coded but directly replaced with the mean value predicted by the hyper-decoder. The skip mode decision is made based on the variance predicted by the hyper-decoder, and when the variance is smaller than a predefined threshold, the skip mode is used. It is assumed that a smaller variance indicates a high confidence in the predicted mean, and the distortion introduced without coding the residual feature is negligible. By contrast, when the skip mode is not used, the codec encodes the residual features element using the mean and variance estimated from the decoded hyperprior model. Different from the skip mode in traditional video compression, where the skip mode is determined by RDO (rate-distortion optimization) through checking all the modes, the skip mode in NVC does not involve an RD cost calculation and comparison process between skip and non-skip modes. Simply comparing the variance with a predefined threshold cannot reflect the RD cost comparison between the different modes. Therefore, to achieve RD optimal coding performance, the RD cost needs to be estimated for the skip and non-skip modes, respectively. 

For the skip mode, no additional rate is incurred in coding the residual feature, and distortion is produced at the reconstruction of the feature. On the contrary, for the non-skip mode, the features are coded with entropy coding, increasing the coding bits, and no distortion is introduced in this entropy coding process. Therefore, the RD cost can be represented as:
\begin{equation}
\label{eq1}
    J = R + \lambda D =
\begin{cases} 
\lambda D_{\text{skip}}, & \text{if skip mode} \\
R_{\text{non-skip}}, & \text{if non-skip mode}
\end{cases}
\end{equation}
where $J$, $R$ and $D$ represent the RD cost, rate and distortion, respectively. The parameter $\lambda$ represents the Lagrange multiplier controlling the rate-distortion trade-off. Here, only the RD cost concerning the skip mode is considered since other rate and distortion do not affect the skip mode decision within the RD cost minimization framework. 

To achieve the RD optimized skip mode decision, the RD cost between the skip and non-skip modes needs to be compared, resulting in the comparison of the rate coding the residual feature and the distortion without coding the feature. Within the NVC framework as shown in Fig.~\ref{structure1}, the entropy model is constructed, usually using a Gaussian based probabilistic model with mean ($\mu$) and variance ($\sigma^2$) estimated by decoding the hyperprior latent variable $\hat{z}$. In this paper, it is assumed that this probabilistic formulation based on the hyperprior follows closely with the actual residual feature, meaning that the actual residual feature also follows the same Gaussian distribution. This is also the case in the current NVC methods when using entropy to approximate the actual bits used to compress the feature. This statistical modeling serves as the basis for formulating the rate and distortion.

For the rate $R_{\text{non-skip}}$, it can be estimated with the entropy:
\begin{equation}
\label{deqn_ex1}
R_{\text{non-skip}}(y) = -\text{log}_2(P(\Delta y|0, \sigma^2I))
\end{equation}
where the probability \( P(\Delta y \mid 0, \sigma^2 I) \) is the centered Gaussian distribution with variance $(\sigma^2)$, $y$ represents the residual feature and $\Delta y =y-\mu$ represents the feature deviation under predicted mean. Note that here, quantization is not considered for simplicity. Otherwise, $y$ represents the quantized residual feature, and $P()$ refers to the cumulative distribution function of the centered Gaussian distribution within the quantization interval.

Since the skip mode decision is based on the predicted variance without the information of the residual feature $y$, for comparing the RD cost, the expected rate $R_{\text{non-skip}}^{E}$ based on the predicted Gaussian distribution is used. 
It can be obtained by integrating over this distribution:
\begin{equation}
\label{rate_E}
\begin{aligned}R_{\text{non-skip}}^E &= \mathbb{E}_{\Delta y : p(y)} \left( R_{\text{non-skip}}(y) \right) \\
&= \int P(\Delta y|0,\sigma^2I)  R_{\text{non-skip}}(y) \, d(\Delta y)  \ \\
&= - \int P(\Delta y|0,\sigma^2I) \text{log}_2(P(\Delta y|0, \sigma^2I)) \, d(\Delta y) 
\end{aligned}
\end{equation}

This formulation estimates the rate, depending only on the  variance $ \sigma $ available at the skip mode decision.

For the distortion estimation $D_{\text{skip}}$, the mean squared error (MSE) in the pixel domain is mostly used. However, the skip mode coding is carried out in the latent space, resulting in distortion in the pixel domain via the decoder. To derive the distortion in the pixel domain from the feature domain, a probabilistic formulation based on Bayes' theorem is constructed. Following the probabilistic model in Ballé et al.~\cite{balle2016end}, the image density conditioned on the latent feature can be represented as $p(x \mid y) = \mathcal{N}(x; \hat{x}, \sigma_x)$, while the latent feature can be represented as $p(y \mid x) = \mathcal{N}(y; \mu, \sigma_y)$. Here, $x$, $\hat{x}$, and $\sigma_x$ represent the image, the reconstructed image (also the mean value in the probabilistic formulation), and the variance of the reconstruction, respectively. $y$, $\mu$, and $\sigma_y$ represent the latent feature, the mean, and the variance of the latent features. The conditional distributions $p(x\mid y)$ and $p(y\mid x)$ correspond to the decoding and encoding processes, respectively. Based on Bayes' theorem, the density function follows:
\begin{equation}
p(x \mid y) = \frac{p(y \mid x) \cdot p(x)}{p(y)}
\end{equation}
where $p(x)$ and $p(y)$ denote the prior distributions of the image x and the latent feature y, respectively, and can be considered independent of the network. By substituting the conditional density function with the assumed Gaussian distribution, the above equation becomes:
\begin{equation}
\begin{aligned}
\frac{1}{\sqrt{2\pi \sigma_x}} \cdot \text{exp}&({\frac{-(x - \hat{x})^2}{2 \sigma_x^2}}) \\
&= \frac{1}{\sqrt{2\pi \sigma_y}} \cdot \text{exp}({\frac{-(y - \mu)^2}{2 \sigma_y^2}}) \cdot \frac{p(x)}{p(y)}
\end{aligned}
\end{equation}
then the following equation can be obtained:  
\begin{align}
\label{distortion_1}
D_{\text{skip}}(y) &= (x - \hat{x})^2 \nonumber \\
&= \frac{\sigma_x^2}{\sigma_y^2} \cdot (y - \mu)^2 + \sigma_x^2 \ln \left( \frac{\sigma_y}{\sigma_x} \right) \nonumber\\
&\quad - 2 \sigma_x^2 \ln \left( \frac{p(x)}{p(y)} \right)
\end{align}

\begin{figure}[!t]
\centering
\includegraphics[width=\columnwidth]{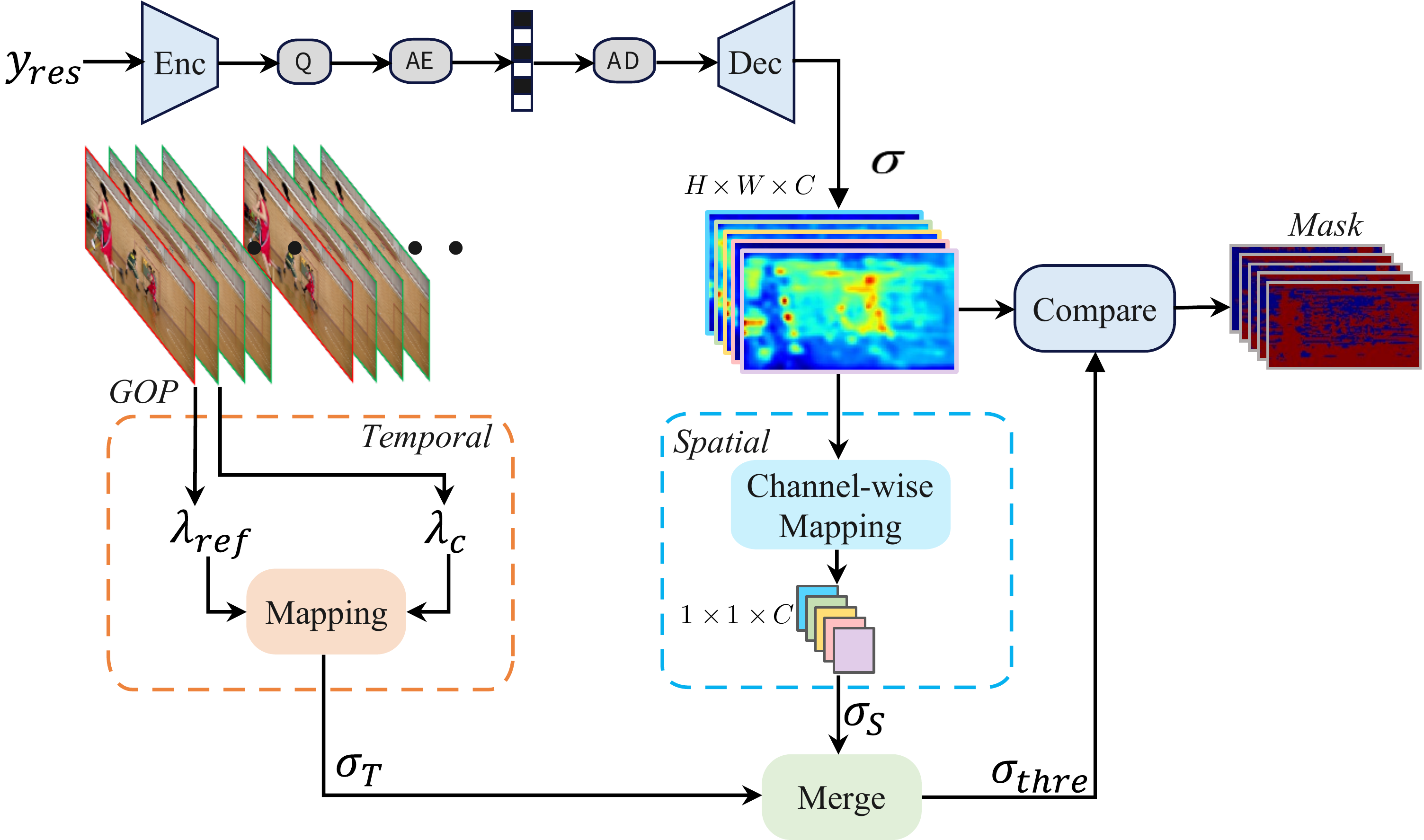}
\caption{Illustrates the architecture of the entropy model integrated with our skip method. $\lambda_{ref}$ and $\lambda_c$ represent the quality of reference frame and current frame, respectively. The spatial module utilizes channel variance to calculate the channel-wise skip threshold, while the temporal method leverages the hierarchical quality structure to apply a temporal threshold.}
\label{structure2}
\end{figure}

Similar to the derivation of the rate term, for skip mode coding, the average distortion over the feature distribution {$P(\Delta y | 0, \sigma^2I)$ } needs to be computed. From this perspective, terms not related to $\Delta y$ can be simplified to a constant since they do not affect the relative change in distortion due to the change in $y$.
Specifically, $p(x)$ and $p(y)$ are prior distributions independent of $\Delta y$, and $\sigma_x$ and $\sigma_y$ are determined by the spatial/temporal context and the network, also independent of $\Delta y$. 
By grouping and simplifying the terms independent of $\Delta y$, Eq. (6) can be further simplified to:
\begin{equation}
\label{distortion_2}
D_{\text{skip}}(y)= J_{\sigma}\cdot(\Delta y)^2+C
\end{equation}
where $\Delta y = y - \mu$ represents the distortion of the feature as in the rate formulation, and $J_\sigma$ and $C$ represent the terms not related to $\Delta y$, $\frac{\sigma_x^2}{\sigma_y^2}$ 
and $\sigma_x^2 \ln \left( \frac{\sigma_y}{\sigma_x} \right) - 2 \sigma_x^2 \ln \left( \frac{p(x)}{p(y)} \right)$ to be specific. This actually agrees with the formulation in AINN\cite{AINN}, where the distortion can be formulated as a combination of the Jacobian matrix of the decoder and the distortion of $y$ under the invertible encoder and the decoder. Accordingly, the expected distortion can be obtained as
\begin{equation}
\label{distortion_3}
\begin{aligned}
D_{\text{skip}}^E &= \mathbb{E}_{y : p(y)} \left( D_{\text{skip}}(y) \right) \\
&= \int P(\Delta y|0,\sigma^2I) D_{\text{skip}}(y) d(\Delta y) \\
&=\int P(\Delta y|0,\sigma^2I) (J_{\sigma}\cdot(\Delta y)^2+C) d(\Delta y) 
\end{aligned}
\end{equation}

By substituting the rate and distortion into Eq.~\ref{eq1}, it can be seen that both the estimated bitrate and the distortion are determined by the hyperprior variance $\sigma$. However, for the rate $R_{\text{non-skip}}^E$, it is solely based on the variance $\sigma$, while for distortion, it is based not only on the variance $\sigma$ but also on the Jacobian matrix $J_\sigma$, which is different for different feature elements. This can also be intuitively understood as the same distortion on different features does not bring same amount of distortion in the reconstruction image. Therefore, the current skip mode decision, based on comparing the variance $\sigma$ with a fixed threshold, cannot achieve the optimal performance since they cannot mimic the RD cost comparison.

To achieve RD optimized skip mode decision, the RD cost of skip mode and non-skip mode need to be compared, which is the comparison of $ R_{\text{non-skip}}^E $ and $ D_{\text{skip}}^E $ as in Eq.~\ref{eq1}. 
However, with the integration operation, it is difficult to directly perform the comparison. Also, computing the Jacobian matrix of the decoder used in the distortion estimation is difficult, which makes element-wise RD optimization for skip mode decision impractical. Therefore, in this paper, we propose a spatially channel-wise RD optimized skip mode decision method based on adaptive variance thresholds.

\begin{figure}[!t]
\centering
\setlength{\belowcaptionskip}{-15pt} 
\includegraphics[width=\columnwidth]{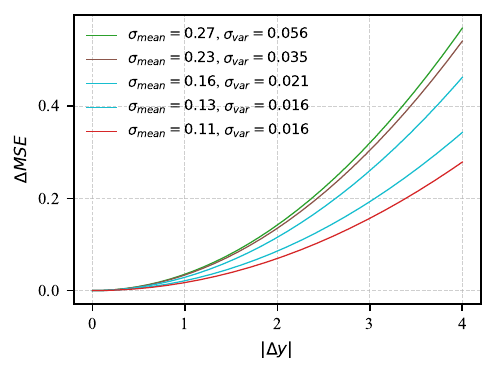}
\caption{Shows the variation of the mean squared error (MSE) with respect to $|\Delta y|$ under different channels, using the scale information for each channel.
}
\label{method1}
 \end{figure}

First, the Jacobian matrix of the decoder is approximated, instead of complicated calculation. Specifically, noise is gradually added into different channels to mimic the skip mode coding process, and the distortion of the reconstructed image corresponding to the feature loss at different channels are shown in Fig.~\ref{method1}, with mean (first order) and variance (second-order) of the variance $ \sigma $ noted for each channel. It can be observed that injecting the same amount of noise into the latent features of different channels leads to different distortion changes of the reconstruction image. Consistent with the derivation in Eq.~\ref{distortion_2}, the variations in $ \Delta y $ are amplified or attenuated by the Jacobian matrix of different channels. For certain channels, the distortion exhibits only small MSE loss with respect to $ \Delta y $, indicating that these channels may carry less information than others. Conversely, other channels are more sensitive to the noise variations in the latent features, indicating that they preserve richer information. 
This demonstrates a clear correlation between the predicted variance statistics and the distortion sensitivity. 
While this cannot provide an analytical solution, it can be empirically used to approximate the distortion change based on the variance statistics.
Moreover, the variance (second-order) of the variance $ \sigma $ within each channel also suggests the feature distribution, e.g., large variance usually suggests rich distribution of the textures. 
Accordingly, the RD optimization of the skip mode decision can be boiled down to the comparison of the variance $ \sigma $ to an adaptive threshold based on the first-order and second-order of the variance $ \sigma $ within the channel. 
In this paper, the adaptive threshold is empirically set to a linear combination $ \sigma_{\text{S}} = \alpha \sigma_{\text{ave}} + \beta \sigma_{\text{std}} + \gamma,$ where $ \sigma_{\text{ave}} $ , $ \sigma_{\text{std}} $ 
represent the mean and standard deviation of the variance $ \sigma $ within each channel, respectively, and $ \alpha $, $ \beta$ and $\gamma$ are parameters empirically set. 
This linear combination of the different-order variance statistics serves as a computationally efficient proxy for Jacobian matrix to approximate the spatial distortion sensitivity. 
It provides a balance between modeling capability and computational efficiency.
In this way, the skip mode can be adaptively determined by taking into consideration the relationship between the distortions of the feature and the reconstructed image.

\begin{table*}[t]
\centering
\captionsetup{justification=centering}
\caption{Performance evaluation (BD-Rate/\%) of the proposed RDOSkip method under AVS CTC, compared with the default skip mode in AVS-EEM}
\label{table1:AVS_CTC}
\begin{tabular}{cc|ccc|ccc|ccc|ccc}
\hline
&\multirow{2}{*}{\textbf{Sequence}} & \multicolumn{3}{c|}{\textbf{Default Skip}} & \multicolumn{3}{c|}{\textbf{w/o Skip}} & \multicolumn{3}{c|}{\textbf{w M8789}} & \multicolumn{3}{c}{\textbf{w RDOSkip}} \\
& & \textbf{Y} & \textbf{U} & \textbf{V} & \textbf{Y} & \textbf{U} & \textbf{V} & \textbf{Y} & \textbf{U} & \textbf{V} & \textbf{Y} & \textbf{U} & \textbf{V} \\
\hline
\multirow{4}{*}{\textbf{HUD4K}} & Tango2&0 &0 &0 & 3.09 & 32.38 & 11.23 & -0.56 & -1.66 & -2.07 & -1.14 & -2.37 & -3.50 \\
&ParkRunning3 &0 &0 &0 & 10.17 & -8.08 & 0.74 & -0.01 & -3.19 & -2.73 & 0.04 & -4.88 & -4.54 \\
&Campfire &0 &0 &0 & 12.73 & -3.9 & 18.62 & -0.35 & 1.8 & -1.2 & -0.57 & 0.60 & -2.82 \\
&DaylightRoad2 &0 &0 &0 & 6.09 & 9.12 & 0.87 & -1.25 & -7.22 & -6.7 & -1.81 & -8.22 & -8.25 \\
\hline
\multirow{4}{*}{\textbf{1080p}} & Cactus &0 &0 &0 & 13.02 & 15.8 & 18.11 & 0.02 & -4.51 & -4.15 & -0.19 & -5.59 & -5.78 \\
&BasketballDrive &0 &0 &0 & 0.29 & 14.37 & 7.39 & -0.91 & -2.2 & -2.29 & -2.10 & -3.70 & -3.93 \\
&MarketPlace &0 &0 &0 & 5.30 & 0.32 & 6.19 & -0.69 & -5.57 & -5.16 & -1.36 & -7.24 & -7.03 \\
&RitualDance &0 &0 &0 & 3.37 & 5.47 & 4.07 & -0.45 & -2.63 & -2.79 & -1.10 & -4.10 & -3.98 \\
\hline
\multirow{4}{*}{\textbf{720p}} & City &0 &0 &0 & 20.25 & 34.27 & 28.38 & -0.53 & -8.24 & -10.13 & -0.77 & -8.72 & -11.19 \\
&Crew &0 &0 &0 & 9.78 & 45.65 & 25.29 & -0.49 & 0.52 & -1.09 & -0.72 & 0.23 & -1.76 \\
&Vidyo1 &0 &0 &0 & 20.44 & 6.96 & 19.36 & -0.7 & -9.22 & -5.45 & -0.70 & -12.00 & -7.49 \\
&Vidyo3 &0 &0 &0 & 19.76 & 24.46 & 21.38 & -0.55 & -8.89 & -7.52 & -0.40 & -10.14 & -7.74 \\
\hline
\hline
\textbf{AVS CTC}& \textbf{Average}&0 &0 &0 & 10.36 & 14.73 & 13.47 & -0.54 & -4.25 & -4.27 & \textbf{-0.90} & \textbf{-5.51} & \textbf{-5.67} \\
\hline
\end{tabular}
\end{table*}

\subsection{Skip Mode Coding Formulation in the Temporal Domain}
\label{3.2}
In the above formulation, the skip mode decision is optimized spatially within each frame based on the rate-distortion cost. In this subsection, the skip mode decision in the temporal domain is investigated. Existing NVC methods, including the AVS-EEM, usually adopt a hierarchical quality structure with different quality parameters \( \lambda \) for a group of frames in the training process. It leads to quality changes among frames, similar to the hierarchical QP setting in traditional video compression. With the context model used in the NVC, the quality of the reference frame may significantly affect the compression of the current frames as shown in Fig.~\ref{intro_3}.

The context-based decoding process in NVC can be represented by:
\begin{equation}
\hat{x}_t = G_s(y_t + \Delta y; \hat{x}_{t-1})
\end{equation}
where \( G_s() \) represents the context-based decoder, \( \hat{x}_t \) and \( \hat{x}_{t-1} \) represent the current reconstructed frame and reference frame, respectively. \( y_t \) and \( \Delta y \) represent the latent feature and distortion of the feature due to skip mode coding, respectively.

Similar to the spatial formulation in Eq.~\ref{distortion_2}, the distortion is controlled by the Jacobian matrix of the decoder. For the context-based decoder, it is related to the reference \( \hat{x}_{t-1} \), which is also the input of the decoder and considered as the condition to the decoding process. Therefore, the hierarchical quality structure also leads to different distributions of the distortion in the reconstruction due to the different qualities of the reference frames. The significantly different distortion distributions under different \( |\Delta y| \) when the reference frame is coded with different qualities shown in Fig.~\ref{intro_3} agree with the above analysis that the quality of the reference frame changes the Jacobian matrix of the context-based decoder.

In contrast, the rate of the latent feature is determined by the entropy model with mean \( \mu \) and variance \( \sigma \) predicted by the hyperprior, not affected by the reference frame. Note that the quality of the reference frame affects the context-based encoding process, thus changing the distribution of the latent feature. However, once the mean \( \mu \) and variance \( \sigma \) are set, the entropy is not affected. In the skip mode decision, the mean and variance are already available, thus not used in the RD cost comparison.

Therefore, a temporally RD optimized skip mode decision method can be further developed considering the change of distortion due to the different qualities of the reference frames. Similarly to the above spatially channel-wise RD optimized skip mode decision, an adaptive threshold is empirically set based on the quality factor, i.e., the Lagrange multiplier \( \lambda \). Together with the spatially channel-wise threshold adaptation, a joint spatial-temporal skip threshold is developed for skip mode coding. Specifically, the adaptive threshold is defined as:
$
\sigma_{\text{Thre}} = \left[ \sigma_{\text{S}} + \varepsilon (\lambda_c - \lambda_{\text{ref}}) \right] \cdot e^{0.3\left(\frac{i}{n} - 1\right)}
$
where \( \lambda_c \) and \( \lambda_{\text{ref}} \) are the Lagrange multipliers of the current frame and its reference frame. \( \varepsilon \) is a factor balancing the Lagrange multiplier change and the variance, which is empirically determined. The $ e^{0.3\left(\frac{i}{n} - 1\right)} $ is a predefined sequence adaptor depending on the index of the current frame, initially set in the AVS-EEM.
 
It is worth noting that this joint spatial-temporal adaptive threshold setting strategy is directly used in the inference without any training. 
On one hand,  {this makes it a lightweight training-free framework without the large computation overhead required for retraining models. }
On the other hand, instead of using learnable parameters, this paper aims to construct a direct and analytical RDO formulation for the skip mode decision and then guide our design. 
It depends on the statistics of the learned model. While it can be adaptively updated in the training process, it may complicate the skip mode decision process.
Overall, the proposed RD optimized skip mode decision method works as a rate distortion optimization process of the entropy mode and can achieve a better trade-off between rate and distortion for skip mode coding.

\section{Experiments}
\subsection{Experimental Setting}

In this paper, the NVC model used in AVS is adopted as our baseline model, AVS-EEM-6.1 to be specific. The model is trained on the Vimeo-90K and BVC-DVC dataset, with the basic $\lambda$ values set to 2048, 512, 256, and 64, to obtain models of four qualities. The GOP-based hierarchical quality setting is used by multiplying an adaptive factor to the Lagrange multipliers $\Delta \lambda$, with GOP size set to 4. Since our method can work directly in the inference without training, in implementing the proposed adaptive skip mode coding, the pre-trained model published by the AVS is adopted. As described in the Introduction, 
the proposed method is only applied to the skip mode coding of the residual information of inter frames, without changing the coding of the motion vector features.
Four hyperparameters, including $\alpha$, $\beta$, $\gamma$, and $\epsilon$, are used in the proposed RDOSkip framework, and are empirically set to $\alpha=0.08$, $\beta=0.015$, $\gamma=0.18$, and $\epsilon=-0.025$, respectively, in the experiments.

\textbf{Test Conditions:} In accordance with the AVS testing, the performance is evaluated under the AVS common test conditions (CTC) \cite{N2654}. 
It includes 12 sequences and all frames are coded with IP (intra period)=-1, meaning only the first frame is coded as intra frame while the other frames are coded as inter frames. Moreover, a GOP size of 4 (GOP=4) is used, where the first frame in a GOP is coded as a high-quality P-frame ($\lambda=1.6$), while the remaining frames are coded as low-quality frames ($\lambda=0.4$). 
Additionally, to validate the effectiveness of our method, experiments on the HEVC dataset \cite{HM}, UVG dataset \cite{UVG}, MCL-JCV dataset \cite{MCL} are conducted. Video sequences of all frames are tested similar to the AVS CTC. Moreover, in addition to the AVS CTC setting, different settings including different intra periods (IP=32), and different GOP sizes (GOP=8) are further evaluated in the ablation study to verify the generalization capability of the proposed method. For IP=32, in every 32 frames, an I frame is inserted. For GOP=8, there is one high-quality P-frame within every 8 frames similarly as in the setting of GOP=4.

In the evaluation, PSRN is used as the quality measure and the BD-rate metric is used as the evaluation criterion, on the Y, U and V channels, respectively.  Four models with different qualities trained with the different Lagrange multipliers are used in order to calculate the BD-Rate over different rates.

\subsection{Performance Comparison}

\begin{table*}[t]
\centering
\captionsetup{justification=centering} 
\caption{Performance evaluation (BD-Rate/\%) of the proposed RDOSkip method under both AVS and HEVC test sequences, compared with the default skip mode in AVS-EEM }
\label{table2:HEVC and CTC}
\begin{tabular}{cc|ccc|ccc|ccc|ccc}
\hline
&\multirow{2}{*}{\textbf{Sequence}} & \multicolumn{3}{c|}{\textbf{Default Skip}} & \multicolumn{3}{c|}{\textbf{w/o Skip}} & \multicolumn{3}{c|}{\textbf{w M8789}} & \multicolumn{3}{c}{\textbf{w RDOSkip}} \\
& & \textbf{Y} & \textbf{U} & \textbf{V} & \textbf{Y} & \textbf{U} & \textbf{V} & \textbf{Y} & \textbf{U} & \textbf{V} & \textbf{Y} & \textbf{U} & \textbf{V} \\
\hline
\multirow{4}{*}{\textbf{AVS CTC}} & HUD4K &0 &0 &0 & 8.02  & 7.38  & 7.86  & -0.54 & -2.57 & -3.18 & -0.98 & -3.20 & -3.56 \\
&1080p &0 &0 &0 & 5.50  & 8.99  & 8.94  & -0.51 & -3.73 & -3.60 & -1.11 & -3.97 & -5.40 \\
&720p &0 &0 &0 & 17.56 & 27.83 & 23.60 & -0.57 & -6.46 & -6.05 & -0.69 & -6.38 & -6.02 \\
\textbf{ }& \textbf{Average}&0 &0 &0 & 10.36 & 14.73 & 13.47 & -0.54 & -4.25 & -4.27& \textbf{-0.90} & \textbf{-5.51} & \textbf{-5.67} \\
\hline
\hline
\multirow{5}{*}{\textbf{HEVC}} & HEVC B &0 &0 &0 & -1.64 & 0.23  & -0.03 & -1.40 & -1.06 & -1.11 & -1.69 & -1.23 & -1.29 \\
&HEVC C &0 &0 &0 & -1.21 & -7.35 & -7.70 & -1.60 & -7.66 & -7.15 & -1.85 & -8.18 & -7.71 \\
&HEVC D &0 &0 &0  & 1.34  & -3.11 & -3.50 & -0.64 & -1.67 & -1.45 & -0.67 & -2.33 & -2.35 \\
&HEVC E &0 &0 &0 & 2.42  & -1.47 & -0.25 & -0.59 & -1.30 & -1.57 & -0.71 & -1.73 & -1.84 \\
\textbf{ }& \textbf{Average} &0 &0 &0 & -0.03& \textbf{-6.19}& \textbf{-6.56}& -1.11 & -4.98 & -4.86 & \textbf{-1.29} & -5.65 & -5.69 \\
\hline
\end{tabular}
\end{table*}

The results of the proposed method under the AVS CTC are shown in Table~\ref{table1:AVS_CTC}, where all frames are encoded with intra period set to -1 (all P frames following the first I frame). The result under ``Default Skip'' represents the AVS codec with the default skip mode coding and serves as the baseline method, while ``w/o Skip'' represents disabling the skip mode for the residual feature coding. ``w M8789'' \cite{M8789} denotes the AVS codec with our temporal adaptive modification method, adopted in AVS. Finally, ``w RDOSkip'' represents the AVS codec with the proposed RD optimized skip mode coding method, which is optimized jointly in both temporal and spatial domains. From Table~\ref{table1:AVS_CTC}, it can be seen that the proposed RDOSkip method consistently achieves better performance than the default skip coding mode, with an average BD-Rate saving of  {0.90} \% on the Y channel. Moreover, it performs better than our temporal skip mode adaptive modification in M8789.

To further demonstrate the effectiveness of the proposed RDOSkip method, performance comparisons on the HEVC test sequences are also evaluated. The results are shown in Table~\ref{table2:HEVC and CTC}. 
It can be seen that the proposed RDOSkip method achieves even better performance with 1.29\% BD-Rate saving on the HEVC test sequences. 
Moreover, by comparing the performance of ``w/o Skip'' with the ''Default Skip'', the default skip mode actually reduces the overall performance for both Class B and C sequences. 
This is mostly because the default skip mode is trained together with the AVS model and may not adapt well to the HEVC test sequences. 
In contrast, the proposed RDOSkip method is not trained and directly used in inference, obtaining consistently better performance than the default skip mode.

\begin{table}[t]
\captionsetup{justification=centering} 
\caption{Percentage of frames with less RD loss than the default skip mode in AVS-EEM under both AVS and \\ HEVC test sequences}
\centering
\label{table3:win rate}
\renewcommand{\arraystretch}{1.2} 
\begin{tabular}{cccccc}
\hline
& \textbf{Sequence} & \textbf{2048} & \textbf{512} & \textbf{128} & \textbf{64} \\ \hline
\multirow{4}{*}{\textbf{AVS CTC}} & HUD4K & 86.96\% & 97.15\% & 97.16\% & 91.15\% \\
& 1080p   & 83.58\% & 95.05\% & 96.63\% & 94.83\% \\
& 720p    & 88.09\% & 92.33\% & 93.25\% & 86.67\% \\
& \textbf{Average}  & 86.21\% & 94.84\% & 95.68\% & 90.88\% \\
\hline
\hline
\multirow{5}{*}{\textbf{HEVC}} &HEVC B & 93.37\% & 94.4\% & 95.88\% & 95.77\% \\
& HEVC C  & 94.46\% & 97.17\% & 98.04\% & 93.84\% \\
& HEVC D  & 79.02\% & 85.87\% & 86.67\% & 84.95\% \\
& HEVC E  & 83.59\% & 90.00\% & 91.84\% & 90.53\% \\ 
& \textbf{Average} & 83.59\% & 90.00\% & 91.84\% & 90.53\% \\
\hline
\end{tabular}
\end{table}

\begin{figure*}[!t]
\centering
\includegraphics[width=1\textwidth]{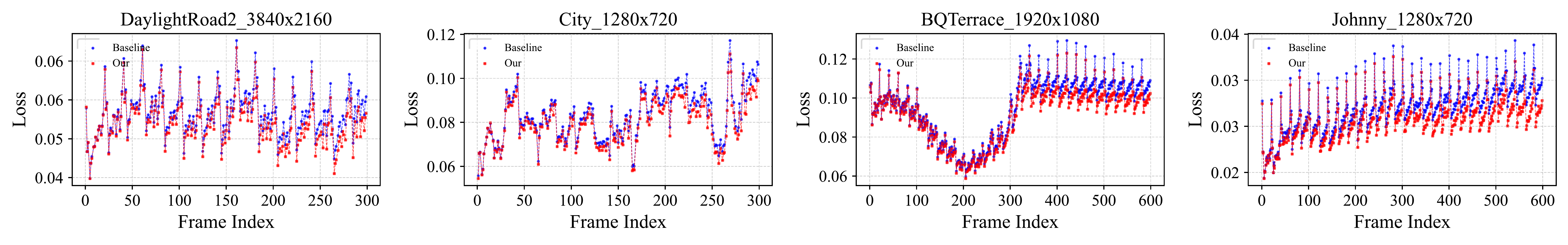}
 \caption{Frame-by-frame loss comparison between the baseline model (the default skip mode in AVS-EEM) and our method. Without high quality P-frame, on AVS CTC dataset(Left) and HEVC dataset(Right), $\lambda=516$. It demonstrates that our approach yields improvements on the majority of frames.}
\label{exp_1}
 \end{figure*}

\begin{figure*}[!t]
\centering
\includegraphics[width=1\textwidth]{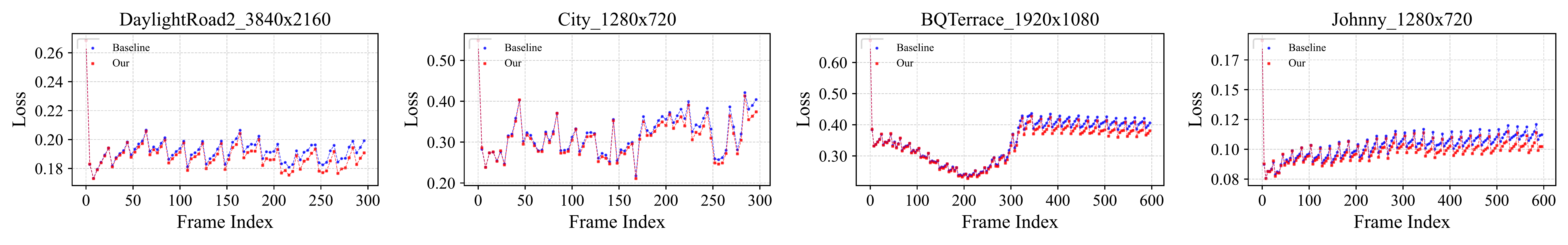}
 \caption{The figure presents the high quality P-frame loss comparison between the baseline model and our method. On AVS CTC dataset(Left) and HEVC dataset(Right), $\lambda=516$. It demonstrates that our approach yields improvements on the majority of high quality frames.}
\label{exp_2}
\end{figure*}

To provide a more comprehensive view, Tables~\ref{table3:win rate} further report the percentages of frames that benefit from our method across different sequences. The RD loss with the Lagrange multiplier used to train each model is calculated for evaluation. The results are illustrated in Table~\ref{table3:win rate}. It can be seen that the proposed method performs better than the baseline skip model on most of the frames, over 90\% on average. The frame-by-frame loss comparison is also illustrated for different sequences as shown in Fig.~\ref{exp_1} and Fig.~\ref{exp_2}. Here, only the results obtained with the model trained under Lagrange multiplier 2048 are used for simplicity, and the results for models under other Lagrange multipliers follow a similar trend. Additionally, due to the use of GOP-based hierarchical quality settings, the RD loss of the first frame in a GOP is much larger than the RD loss of the following three frames due to the large Lagrange multiplier used for the first frame. Therefore, for better illustration, RD loss comparisons for the first frame and the following three frames (with the same Lagrange multiplier) are shown separately. It can be seen that our method achieves smaller RD loss per-frame than the default skip model. Especially on the HEVC Class B, C dataset, the performance improvement margin is large, which also agrees with the BD-Rate saving as shown in Table~\ref{table2:HEVC and CTC}.

\begin{table}[t]
\centering
\captionsetup{justification=centering} 
\caption{Performance evaluation (BD-Rate/\%) of different modules in the proposed method under both AVS and HEVC test sequences on the Y channel}
\label{table4:ablation1}
\renewcommand{\arraystretch}{1.2} 
\begin{tabular}{ccccc}
\hline
&\textbf{Model} & \textbf{T-skip} & \textbf{S-skip} & \textbf{BD-Rate (\%)} \\ 
\hline
\multirow{3}{*}{\textbf{AVS CTC}} & M0 & $\times$ & $\times$ & 0 \\
&M1 & $\checkmark$ & $\times$ & -0.54 \\
&M2 & $\checkmark$ & $\checkmark$ & -0.90\\ 
\hline
\multirow{3}{*}{\textbf{HEVC}} & M0 & $\times$ & $\times$ & 0 \\
&M1 & $\checkmark$ & $\times$ & -1.11 \\
&M2 & $\checkmark$ & $\checkmark$ & -1.29 \\ 
\hline
\end{tabular}
\end{table}

\subsection{Ablation Studies}
Extensive ablation experiments are conducted to verify the effectiveness of the proposed method, including evaluation on the different modules, different GOP sizes, different intra periods. Visual analysis and complexity evaluation is also conducted and illustrated in this subsection. Additionally, experiments on more datasets including the UVG and the MCL-JCV datasets, on more NVC backbones including the DCVC-RT \cite{DCVC-RT}, and the parameter sensitivity analysis are further conducted and described in the supplementary material.

\textbf{Evaluation on different modules:} To verify the effectiveness of the proposed spatial and temporal skip mode coding methods, an ablation study on the different modules is conducted. AVS-EEM with the default skip mode is used as baseline (M0) when calculating the BD-rate. Then the proposed modules are progressively added on top of the baseline. The results are shown in Table~\ref{table4:ablation1}. It can be seen that the proposed temporal adaptive skip mode coding method (noted as ``T-skip'' in the table) can bring 0.54\% and 1.11\% BD-Rate reduction on AVS CTC dataset and HEVC dataset, respectively. Then the spatial adaptive skip mode coding method  (noted as ``S-skip'' in the table) further improves the performance to   {0.90} \% and 1.29\% BD-Rate saving on AVS CTC dateset and HEVC dataset, respectively, demonstrating the effectiveness of the proposed joint spatial-temporal adaptive skip mode coding. 

\textbf{Evaluation on different GOP sizes:} In view that our temporal adaptive skip coding method is related to the structure of the frame qualities in the GOP setting, the proposed method under different group sizes are also evaluated. In addition to the setting of GOP = 4, the GOP setting of GOP = 8 is tested, where there is one high-quality P-frame within every 8 frames. The results are shown in Table~\ref{table5:ablation_GOP}. It can be seen that the proposed method also improves the performance by 0.68\% and 0.85\% BD-Rate saving on the Y channel. Note that here we used the same set of parameters as in GOP=4 for the test. This further demonstrates that our temporal adaptive skip mode coding method can be very robust to different GOP settings.

\begin{table}[t]
 \captionsetup{justification=centering} 
 \centering
\caption{Performance evaluation (BD-Rate/\%) of different modules in the proposed method under different GOP settings on the Y channel}
\label{table5:ablation_GOP}
\renewcommand{\arraystretch}{1.2} 
\begin{tabular}{ccccc}
\hline
&\textbf{Model} & \textbf{T-skip} & \textbf{S-skip} & \textbf{BD-Rate (\%)} \\ 
\hline
\multirow{3}{*}{\textbf{AVS CTC}} & M0 & $\times$ & $\times$ & 0 \\
&M1 & $\checkmark$ & $\times$ & -0.36 \\
&M2 & $\checkmark$ & $\checkmark$ & -0.68\\ 
\hline
\multirow{3}{*}{\textbf{HEVC}} & M0 & $\times$ & $\times$ & 0 \\
&M1 & $\checkmark$ & $\times$ & -0.82 \\
&M2 & $\checkmark$ & $\checkmark$ & -0.85 \\ 
\hline
\end{tabular}
\end{table}

\textbf{Evaluation on different intra periods:} In addition to all P frame used in the AVS CTC, the setting of intra period (IP) equaling 32 (IP=32) is also evaluated. In this case, the threshold decay $e^{0.3(\frac{i}{n}-1)}$ in the default skip mode is also changed to $e^{0.3(\frac{i}{32}-1)}$, to accommodate the change of intra period. Here $i$ and $n$ represent the current frame index and the total frame number, respectively. It works as a temporal decay weight to improve the overall performance. When IP=32, the video sequence inserts an I frame for each 32 frames and separates the sequence into different groups, thus changing $n$ to the intra period size. The results are shown in Table~\ref{table6:IP32}. It is known that the bit-rate of encoding an I frame is much larger than that of encoding a P frame, and when the number of P-frames is relatively small under small intra period, the rate of I-frames can affect the overall performance relatively large. Therefore, the performance is only calculated for the P-frames to better illustrate the performance change since our method is only applied to the P-frames. From the results shown in Table\ref{table6:IP32}, it can be seen that our method also improves the performance by 1.62\% Y channel BD-Rate saving over the default skip mode in AVS-EEM.

\begin{figure*}[!t]
\centering
\includegraphics[width=1\textwidth]{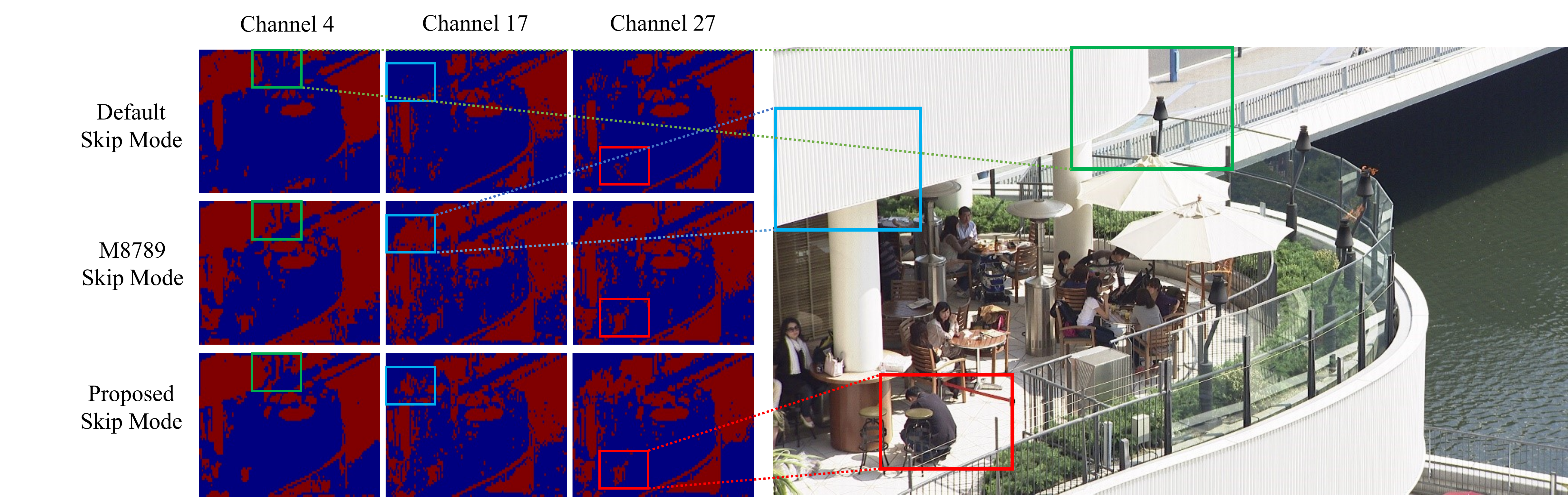}
\caption{Example skip mode masks using different methods, evaluated on the $2^{nd}$ frame of the sequence BQTerrace. Red points indicates elements skipped while blue points indicates those non-skipped.}
\label{exp_3}
\end{figure*}

\begin{table*}[!t]
\centering
\captionsetup{justification=centering} 
\caption{Performance evaluation (BD-Rate/\%) of the proposed RDOSkip method under both AVS and HEVC test sequences, with intra period setting IP=32, compared with the default skip mode in AVS-EEM }
\label{table6:IP32}
\begin{tabular}{cc|ccc|ccc|ccc|ccc}
\hline
&\multirow{2}{*}{\textbf{Sequence}} & \multicolumn{3}{c|}{\textbf{Default Skip}} & \multicolumn{3}{c|}{\textbf{w/o Skip}} & \multicolumn{3}{c|}{\textbf{w M8789}} & \multicolumn{3}{c}{\textbf{w RDOSkip}} \\
 & & Y& U& V& Y& U& V& Y& U& V& Y& U&V\\
\hline
\multirow{4}{*}{\text{AVS CTC}} & HUD4K &0 &0 &0 & 6.38  & -2.94& -0.42& -0.18 & 0.73  & 0.45  & -0.85 & 1.55  & 1.35 \\
& 1080p &0 &0 &0 & 3.17  & 0.84  & 0.02  & -0.19 & -0.20 & 0.31  & -0.76 & -0.27 & 0.16 \\
& 720p  &0 &0 &0 & 9.31  & 10.80 & 8.38  & -0.74 & -0.89 & -0.96 & -2.75 & -1.60 & -2.01 \\
\textbf{ }& \textbf{Average} &0 &0 &0 & 6.29 & 2.90 & 2.66  & -0.37 & -0.12 & -0.07 & \textbf{-1.45} & \textbf{-0.11} & \textbf{-0.17} \\
\hline
\hline
\multirow{5}{*}{\text{HEVC}} & HEVC B &0 &0 &0 & 4.16  & 5.34  & 5.14  & -0.85& -1.18 & -1.12 & -1.25 & -1.75& -1.67 \\
& HEVC C &0 &0 &0 & 4.77  & 4.21  & 4.24  & -1.33 & -0.92& -0.86 & -2.1& -2.32& -2.53\\
& HEVC D &0 &0 &0 & 5.08  & 2.86  & 1.87  & -0.8& 0.63& 1.24& -1.03& -2.38& -1.64\\
& HEVC E &0 &0 &0 & 7.34  & 6.17  & 6.33  & -1.73& -0.87& -0.67& -2.64 & -1.49& -1.19\\
\textbf{ }& \textbf{Average} &0 &0 &0 & 5.14  & 3.24  & 2.97  & -1.12 & -0.34 & -0.14 & \textbf{-1.67} & \textbf{-1.64} & \textbf{-1.47} \\
\hline
\end{tabular}
\end{table*}

\textbf{Visualization of the proposed method:} To provide an intuitive understanding of the proposed adaptive skip mode coding module, Fig.~\ref{exp_3} presents the impact of the proposed method on the spatial channel-wise masks. It can be seen that our method is able to mask out certain less important regions. The red regions in Fig.~\ref{exp_3} indicate the skipped areas. It can be observed that, in the default skip mode, some smooth background regions are not skipped, leading to higher bitstream consumption (e.g., in the green and blue boxes). With our adaptive skip mode, such regions can be skipped. Certain important features can also be preserved in our method as shown in the red box in Fig.~\ref{exp_3}, thereby retaining more fine structural details.


\textbf{Evaluation on the complexity:} The complexity of the proposed method is negligible, with no learnable parameter increase and almost unnoticeable computation increase. 
Since it is only applied in the test, it does not add any training cost. 
The encoding and decoding speed in terms of FPS (frames per second) is evaluated on the AVS CTC dataset, and the results are shown in Table~\ref{tab:runtime_avs}. The test is performed on an NVIDIA V100 GPU. It can be seen that the proposed method shows no complexity increase in terms of encoding and decoding time, and even slightly improves the speed. 
This is mostly because with our RDOSkip, more feature elements may be skipped, thus reducing the following entropy coding complexity. 

\begin{table}[t]
  \centering
  \captionsetup{justification=centering}
  \caption{Encoding and decoding complexity evaluation in terms of FPS (frames per second) on the AVS CTC Dataset in AVS-EEM}
  \label{tab:runtime_avs}
  \begin{tabular}{lcccc}
    \toprule
    \multirow{2}{*}{\textbf{Dataset}} & \multicolumn{2}{c}{\textbf{Encoding (FPS)}} & \multicolumn{2}{c}{\textbf{Decoding (FPS)}} \\
    \cmidrule(lr){2-3} \cmidrule(lr){4-5}
    & \textbf{Baseline} & \textbf{RDOSkip} & \textbf{Baseline} & \textbf{RDOSkip} \\
    \midrule
    AVS CTC 4K & 2.95  & 2.94  & 2.89  & 2.99  \\
    AVS CTC 1080p & 11.47 & 11.53 & 9.94 & 9.59 \\
    AVS CTC 720p  & 18.90 & 19.14 & 12.40 & 12.80 \\
    \bottomrule
  \end{tabular}
\end{table}

\section{Conclusion}
This paper proposes a rate-distortion optimized skip mode coding (RDOSkip) framework for NVC. It first formulates the rate and distortion in NVC, in order to realize RD cost comparison between skip mode and non-skip mode. Based on the RD formulation, a spatially channel-wise RD optimized skip mode decision method is developed to facilitate the adaptive skip mode coding among channels within a spatial frame. Considering the GOP based temporal hierarchal quality structure, a temporally RD optimized skip mode decision method is further developed to accommodate the distortion change caused by the context based decoder and reference frame quality fluctuation. Extensive experiments are conducted on the AVS NVC reference software and the results demonstrate that our method improves the performance by 0.90\% on average in terms of BD-rate saving without any training.

\bibliographystyle{IEEEtran} 
\bibliography{references}


\begin{IEEEbiography}[{\includegraphics[width=1in,clip,keepaspectratio]{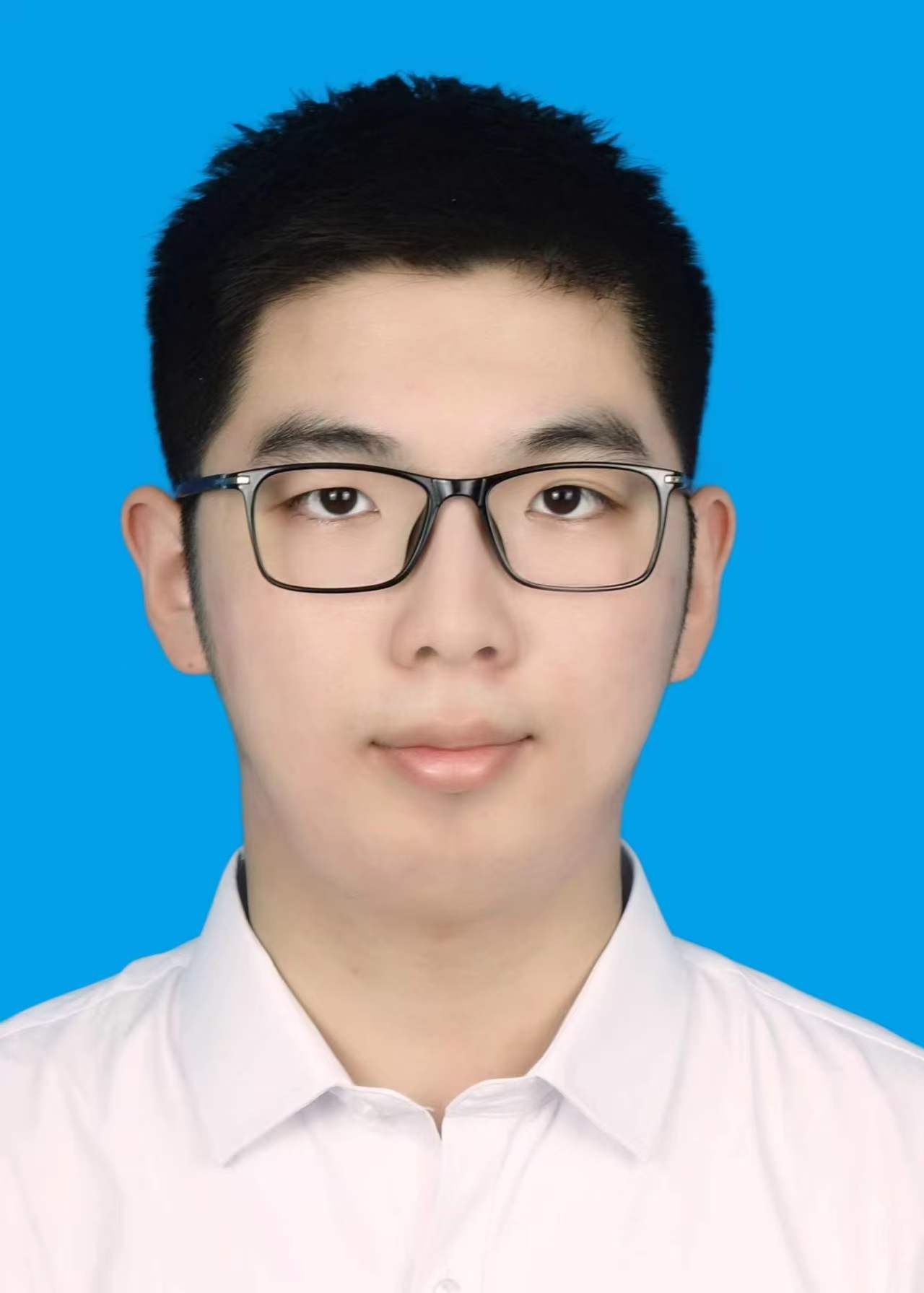}}]{Yuntian Jiang}
received B.S. degree in Robotics Engineering from University of Shanghai for Science and Technology, Shanghai, China, in 2024. He is currently pursuing the M.S. degree with the School of Control Science and Engineering in Shandong University. His research interests include video compression and deep learning. 
\end{IEEEbiography}

\begin{IEEEbiography}[{\includegraphics[width=1in,clip,keepaspectratio]{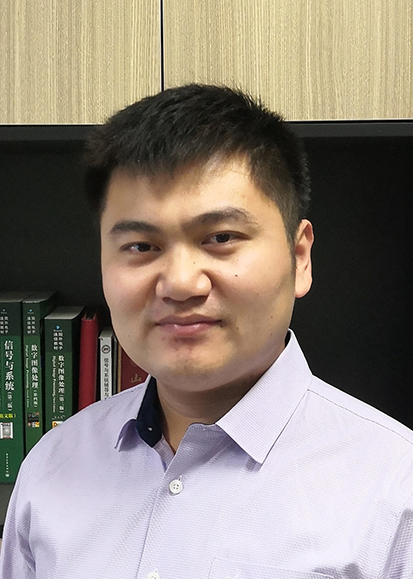}}]{Shuai Li}
 is currently with the School of Control Science and Engineering, ShanDong University (SDU), China, as a Professor and QiLu Young Scholar. He was with the School of Information and Communication Engineering, University of Electronic Science and Technology of China, China, as an Associate Professor from 2018-2020. He received his Ph.D. degree from the University of Wollongong, Australia, in 2018. His research interests include image/video coding, 3D video processing and computer vision. He was a co-recipient of two best paper awards at the IEEE BMSB 2014 and IIH-MSP 2013, respectively.
\end{IEEEbiography}

\begin{IEEEbiography}[{\includegraphics[width=1in,clip,keepaspectratio]{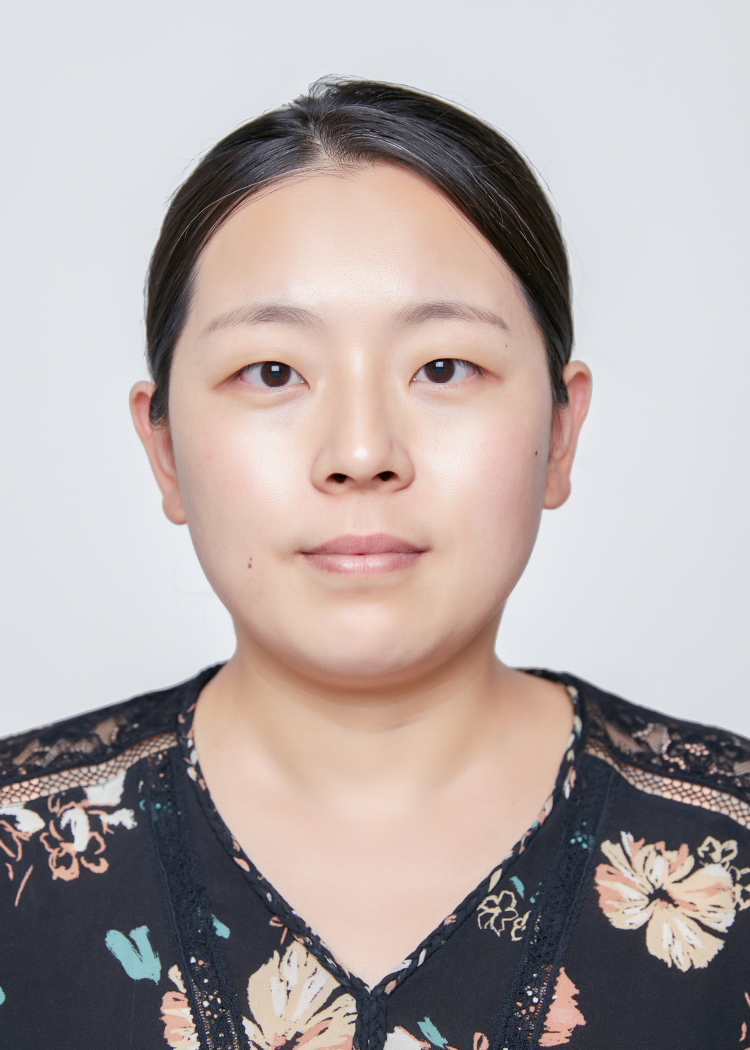}}]{Yanbo Gao}
 is currently with the School of Software, Shandong University (SDU), Jinan, China, as an Associate Professor. She was with the School of Information and Communication Engineering, University of Electronic Science and Technology of China (UESTC), Chengdu, China, as a Post-doctor from 2018-2020. She received her Ph.D. degree from UESTC in 2018. Her research interests include video coding and 3D video processing. She was a co-recipient of the best student paper awards at the IEEE BMSB 2018.
\end{IEEEbiography}

\begin{IEEEbiography}[{\includegraphics[width=1in,clip,keepaspectratio]{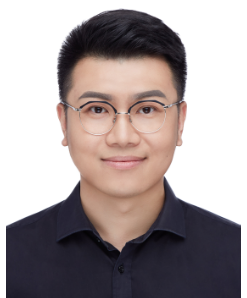}}]{Jinglin Zhang}
  received the B.E. degree in communication engineering from South Central University
  for Nationalities, Wuhan, China, in 2007, the
  M.E. degree in electronic circuit and system from
  Shanghai University, Shanghai, China, in 2010, and
  the Ph.D. degree in electronics and communication
  engineering from the National Institute of Applied
  Sciences, Rennes, France, in 2013.
  He is currently a Professor with the School of Control Science and Engineering, Shandong University,
  Jinan, China. His research interests include computer
  vision and interdisciplinary research with pattern recognition and atmospheric
  science.
\end{IEEEbiography}

\begin{IEEEbiography}[{\includegraphics[width=1in,clip,keepaspectratio]{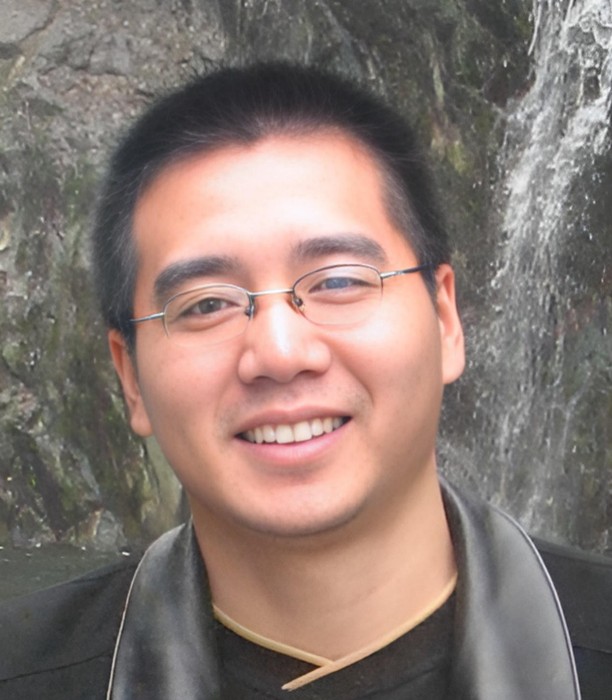}}]{Hui Yuan} 
 (S’08–M’12–SM’17) received the B.E. and Ph.D. degrees in telecommunication engineering from Xidian University, Xi’an, China, in 2006 and 2011, respectively. In April 2011, he joined Shandong University, Ji’nan, China, as a Lecturer (April 2011–December 2014), an Associate Professor (January 2015-October 2016), and a Professor (September 2016). From January 2013-December 2014, and November 2017-February 2018, he also worked as a Postdoctoral Fellow (Granted by the Hong Kong Scholar Project) and a Research Fellow, respectively, with the Department of Computer Science, City University of Hong Kong, Hong Kong. From November 2020 to November 2021, he also worked as a Marie Curie Fellow (Granted by the Marie Skłodowska-Curie Individual Fellowships of European Commission) with the Faculty of Computing, Engineering and Media, De Montfort University, United Kingdom. From October 2021 to November 2021, he also worked as a visiting researcher (secondment of the Marie Skłodowska-Curie Individual Fellowships) with the Computer Vision and Graphics group, Fraunhofer Heinrich-Hertz-Institut (HHI), Germany. His current research interests include 3D visual coding, processing, and communication. He is serving as an Associate Editor for IEEE Transactions on Image Processing and IEEE Transactions on Consumer Electronics.
\end{IEEEbiography}
 \vspace{-0.8cm}
\begin{IEEEbiography}[{\includegraphics[width=1in,clip,keepaspectratio]{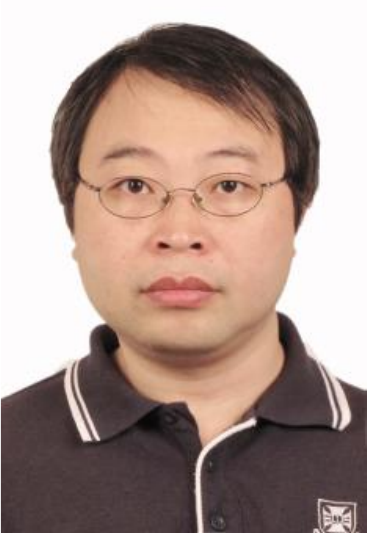}}]{Mao Ye}	 (Senior Member, IEEE) received the B.S. degree from Sichuan Normal University, Chengdu, China, in 1995, the M.S. degree from the University of Electronic Science and Technology of China, Chengdu, in 1998, and the Ph.D. degree from The Chinese University of Hong Kong, Hong Kong, in 2002, all in mathematics. He has been a short-time Visiting Scholar at the University of Queensland and the University of Pennsylvania. He is currently a Professor and the director of CVLab, University of Electronic Science and Technology of China, Chengdu. His research interests include machine learning and computer vision. In these areas, he has published over 90 articles in leading international journals or conference proceedings. He was a co-recipient of the Best Student Paper Award at the IEEE ICME 2017. He has served on the Editorial Board of the Engineering Applications of Artificial Intelligence.
\end{IEEEbiography}

\vfill

\end{document}